\documentclass[10pt,A4paper]{article}
\usepackage[top=0.85in,left=2.75in,footskip=0.75in,marginparwidth=0in]{geometry}

\usepackage[utf8]{inputenc}

\usepackage{cite}

\usepackage{nameref,hyperref}

\usepackage[right]{lineno}

\usepackage{microtype}
\DisableLigatures[f]{encoding = *, family = * }

\raggedright
\usepackage{changepage}

\usepackage[aboveskip=1pt,labelfont=bf,labelsep=period,singlelinecheck=off]{caption}

\makeatletter
\renewcommand{\@biblabel}[1]{\quad#1.}
\makeatother

\usepackage{lastpage,fancyhdr,graphicx}
\usepackage{epstopdf}

\usepackage{amsbsy,amssymb,amsmath,bm,bbold}
\usepackage{graphicx,color,epsfig,rotate} 
\usepackage{fancyhdr} 
\usepackage{epstopdf}
\usepackage{float}
\usepackage{csquotes}
\usepackage[justification=justified]{caption}
\usepackage{multirow}
\usepackage{booktabs,makecell}

\usepackage{siunitx}
\usepackage{ragged2e}
 
\fancyheadoffset[L]{2.25in}
\fancyfootoffset[L]{2.25in}

\usepackage{color}

\definecolor{Gray}{gray}{.25}

\usepackage{graphicx}

\usepackage{sidecap}

\usepackage{wrapfig}
\usepackage[pscoord]{eso-pic}
\usepackage[fulladjust]{marginnote}
\reversemarginpar

\begin{document}
\vspace*{0.35in}

% title goes here:
\begin{center}
{\Large
\textbf\newline{Unconventional next nearest neighbor resonance coupling and states flipping mechanism in degenerate optical microcavities}
}
\newline
% authors go here:
\\
Arnab Laha\textsuperscript{1},
Abhijit Biswas\textsuperscript{2},
Somnath Ghosh\textsuperscript{1,*}
\\
\bigskip
\bf{1} Department of Physics, Indian Institute of Technology Jodhpur, Rajasthan 342011, India
\\
\bf{2} Institute of Radio Physics and Electronics, University of Calcutta, Kolkata-700009, India
\\
\bigskip
* somiit@rediffmail.com

\end{center}

\justify
\section*{Abstract}
We report a specially configured non-Hermitian optical microcavity, imposing spatially imbalanced gain-loss profile, to host an exclusively proposed next nearest neighbor resonances coupling scheme. Adopting scattering matrix ($S$-matrix) formalism, the effect of interplay between such proposed resonance interactions and the incorporated non-Hermiticity in the microcavity is analyzed drawing a special attention to the existence of hidden singularities, namely exceptional points ($EP$s); where at least two coupled resonances coalesce. We establish adiabatic flip-of-states phenomena of the coupled resonances in the complex frequency plane ($k$-plane) which is essentially an outcome of the fact that the respective $EP$ is being encircled in system parameter plane. Encountering such multiple $EP$s, the robustness of flip-of-states phenomena have been analyzed via continuous tuning of coupling parameters along a special hidden singular line which connects all the $EP$s in the cavity. Such a numerically devised cavity, incorporating the exclusive next neighbor coupling scheme, have been designed for the first time to study the unconventional optical phenomena in the vicinity of $EP$s.

% now start line numbers
%\linenumbers

% the * after section prevents numbering
\section{Introduction}
Over the years, resonance interaction phenomena in open quantum systems have been attracted enormous attention in various research field of modern physics. Various interesting interaction phenomena exploiting local and non-local interdependence between the resonance states have reported in literature. Specifically, in the photonics domain, interesting techniques have developed for modeling and simulation of different specially configured coupled optical systems to study such interactions between the states. This paper present a specially configured coupled optical system with discrete resonances where interesting effects of next nearest neighbor interaction between them have been topologically explored. In the contemporary research field, next nearest neighbor interaction between the resonances has always been a great physical insight because it is a pivotal feature in many natural and artificial physical phenomena. Statistically, 1D Ising model, a mathematical model of ferromagnetism in solid state physics, gives a clear interpretations of next nearest neighbor interaction, while considering the physical effect of superimposition of very long range spin interaction with conventional nearest neighbor short range interaction on a 1D crystalline lattice~\cite{Kijewski}. Lately, next nearest neighbor interactions have also been explored in the contexts of QCD through three states Potts model (a generalization of the Ising model)~\cite{Bernaschi}, Betts lattice considering extended Hubbard model to study pairing enhancement~\cite{Fang}. Influence of such interactions phenomena have also attracted considerable attention to study various physical applications like entanglement of the Heisenberg chain~\cite{Gu}, thermal transportation in low dimensional lattice~\cite{Santhosh}, spectrum of plasmon excitations in graphene (considering next-nearest-neighbor tight-binding model)~\cite{Kadirko} etc. In the optical context lately, effect of next nearest neighbor coupling have widely discussed on optically pumped nanodevice arrays~\cite{Csaba}, Bose–Einstein condensation in optical lattices~\cite{Zaleski}, photonic superlattice to implements 1D random mass Dirac equation on a chip~\cite{Keil} etc.

Apart from the previous studies, corroborating the analogy between non-Hermitian open quantum system and counterpart open optical geometries with suitable pumping, we explore an innovative unconventional scheme to study a nontrivial special next nearest neighbor interaction between discrete resonance states in a coupled optical microcavity. The cavity is partially pumped via spatially distributed inhomogeneous gain-loss profile. In such a cavity the resonances are appeared in complex energy plane. Here the coupling phenomena between the resonances are entirely controlled internally i.e. by system topology and internal gain-loss variation.
\begin{figure}[t]
	\centering
	\includegraphics[width=7cm]{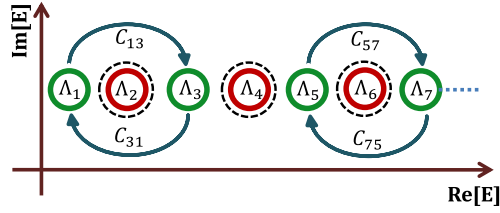}
	\vspace{0.2 cm}
	\caption{(Color online) Schematic diagram of the proposed coupling scheme between the resonances appeared in complex energy plane. The resonances labeled by green colors are being interacted whereas the resonances labeled by red colors remain isolated.   }
	\label{figure_1}
\end{figure}      

The special non-trivial coupling scheme, schematically shown in figure~\ref{figure_1}, is proposed in such a way that, a specific resonance is strictly allowed to interact with its next nearest neighbor only with a restriction of one-to-one coupling when the intermediate resonances between each of the two coupled states remains isolated. Proposed coupling scheme can be mathematically demonstrated by the following Hamiltonian function described below.

Consider a quantum mechanical coupled system subjected by an external field $h_n$ with discrete resonances characterized by the parameters $\Lambda_n (n=1,2,3...)$. Now the Hamiltonian can be written as-
\begin{equation}
{H}=p\sum\limits_{i,j} C_{2i-1,2j+1}\Lambda_{2i-1}\Lambda_{2j+1} +q\sum\limits_{k,l} C_{2k-1,2l+1}\Lambda_{2k-1}\Lambda_{2l+1}+r\sum\limits_{n}h_n\Lambda_n
\label{equation_1}  
\end{equation}
Here, $i,j=1,3,5....2n+1$ and $k,l=2,4,6....2n$ $(i,j,k,l\in n)$.
The coefficients $C$ indicate the interaction of one resonance to its next nearest neighbor. $p$, $q$ and $r$ are the real dimensionless parameters. For our proposed scheme, there must be one isolated resonance between two coupled resonances. i.e. the resonances appears in sites labeled by odd integers ($i$ and $j$) are interacting, whereas the resonances appears in sites labeled by even integers ($k$ and $l$) remain isolated. So, purposely choosing the parameter $q$ as 0 we neglect the second term of Eqn.~\ref{equation_1}. Here we also deliberately neglect all possible coupling phenomena of the resonances with external field and henceforth set the parameter $r$ at 0. So, according to the proposed coupling scheme the Hamiltonian function (Eqn.~\ref{equation_1}) must be reformed as  
\begin{equation}
{H}=p\sum\limits_{i,j} C_{2i-1,2j+1}\Lambda_{2i-1}\Lambda_{2j+1};\quad i,j=1,3,5.. 
\label{equation_2}  
\end{equation}

Towards the topological studies based on resonance interactions in such a non-Hermitian open optical microcavity the phenomena of avoided resonance crossing ($ARC$) play a key role~\cite{Heiss1,Cartarius3,Laha2,Ghosh,Laha1}. Usually, $ARC$ occur in complex energy plane where two interacting resonances repel each other via crossing/anticrossing of their energies and widths i.e. essentially their real and imaginary parts. Such $ARC$ phenomena between two interacting resonance states  have been referred the presence of a specific spectral singularity where they are very close to a special type of degeneracy which is rather different form genuine Hermitian degeneracy. These specific hidden spectral singularities, usually appeared in parameter space with at least either two real valued parameters or a complex parameter, are named as {\it exceptional points} ($EP$s). At these $EP$s the system Hamiltonian becomes defective and two coupled levels coalesce~\cite{Heiss1,Cartarius3,Laha2,Ghosh}. 

An $EP$ leads to crucial modifications on associated coupled eigenvalues' behavior under the influence of coupling parameters; where the phenomenon of flipping of states in the complex eigenvalue plane is the most significant in the context of optical mode converter~\cite{Laha2,Ghosh}. In parameter space, adiabatically a moderate variation of the chosen coupling parameters along a closed contour around an $EP$ results in the permutation between the corresponding coupled eigenvalues (exchanging their positions) in complex energy plane exhibiting $EP$ as a second order branch point~\cite{Cartarius3,Laha2,Ghosh,Menke}. Consequently, the corresponding eigenstates are also permuted exhibiting $EP$ as a forth order branch point followed by an additional phase change after each round~\cite{Heiss1} in a manner like $\{\psi_1,\psi_2\}\rightarrow\{\psi_2,-\psi_1\}$. By contrast, for an $EP$ which is not being enclosed by the parametric contour; associated eigenvalues make individual loop and avoid any kind of permutation. This unique features of {\it flip-of-state} phenomenon in the vicinity of $EP$s have theoretically been explored in various non-Hermitian systems like atomic~\cite{Cartarius3,Menke} as well as molecular~\cite{Lefebvre} spectra, partially pumped optical microcavity~\cite{Laha2}, laser~\cite{Berry}, optical waveguide~\cite{Ghosh} etc. and also verified experimentally~\cite{Dembowski2}. Technologically, this unconventional phenomena leads a key feature towards sensor operation~\cite{Wiersig1} in the context of single particle detection in microcavity~\cite{Wiersig2} and also for mode management in dark-state laser~\cite{Hodaei}.

In this paper for the first time to the best of our knowledge, we explore $EP$s with their unconventional specific features in a non-Hermitian optical microcavity operating under the proposed non-trivial next-nearest-neighbor resonance coupling condition. A specially configured non $\mathcal{PT}$-symmetric Fabry-Perot type optical microcavity with spatially unbalanced gain-loss profile has been reported for this specific purpose. Recent advanced development in fabrication technology for growth of such Fabry-Perot type optical micro-resonators with enhanced precision and control on output coupling without proper phase matching (which is not possible for other geometries) have resulted in extensive contemporary research attention towards easier technological implementation. Numerically designing such a microcavity, we encounter at least three second order $EP$s in the functional parameter space of the cavity via internally coupled resonances situated in next nearest neighbor positions with rigid one-to-one coupling restriction which is entirely controlled by topology of the microcavity. We also establish a formation of special hidden singular line, namely {\it exceptional line}, to correlate the identified $EP$s in parameter plane. Very recently such correlation have reported by the authors in the similar form of the optical microcavity operating under nearest neighbor resonance interaction only~\cite{Laha2}. Unconventional cascaded flip-of-state mechanism in complex frequency plane ($k$-plane) with its robustness against parameter fluctuations/ deformations has also reported in the vicinity of $EP$s by encircling multiple $EP$s in parameter plane via continuous tuning of coupling parameters along exceptional line. Overall, exploiting $EP$s the optical performances of the microcavity have judicially tailored under restricted operating condition.

\section{Matrix formulation towards appearance of exceptional points incorporating next nearest neighbor resonance coupling} \label{matrix_formulation}

Mathematically, to elaborate all the essential aspects of a second order $EP$ under specially proposed next nearest neighbor coupling scheme as described by Eqn.~\ref{equation_2}, a matrix formulation should be needed. The proposed scheme demands at least a real $3 \times 3$ Hamiltonian for this specific purpose. The following Hamiltonian $H$ with the form $H_0+\lambda H_p$  describes a passive quantum system $H_0$ with three discrete energy eigenvalues $\varepsilon_i (i = 1,2,3)$, which is subjected by a perturbation $H_p$. i.e.
\begin{equation}
{H}=\left(\begin{array}{ccc}\varepsilon_1 & 0 & 0 \\0 & \varepsilon_2 & 0 \\0 & 0 & \varepsilon_3\end {array}\right)+\lambda U\left(\begin{array}{ccc}\omega_1 & 0 & 0 \\0 & \omega_2\ & 0 \\ 0 & 0 & \omega_3\end {array}\right)U^\dagger
\label{equation_3}  
\end{equation}
where,
\begin{equation}
{U(\xi)}=\left(\begin{array}{ccc}\cos\xi & 0 & -\sin\xi \\0 & 1 & 0 \\\sin\xi & 0 & \cos\xi\end {array}\right)
\label{equation_4}  
\end{equation}

\begin{figure}[t]
	\centering
	\includegraphics[trim={0.7cm 1.5cm 1cm 0cm},width=8.5cm]{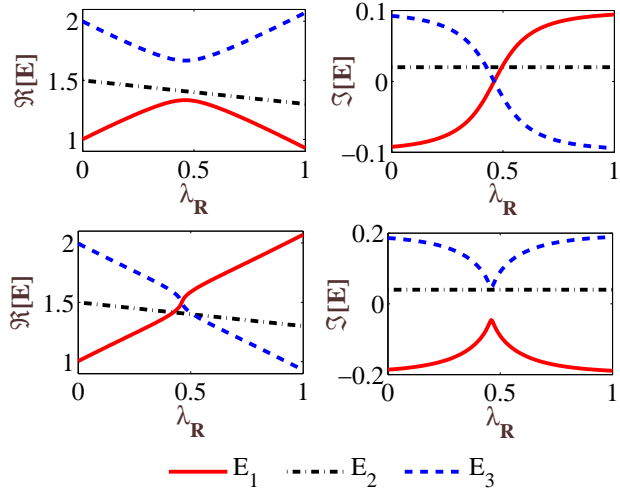}
	\vspace{0.2 cm}
	\caption{(Color online) Level repulsion phenomena with variation of $\lambda$ showing crossing/ anticrossing of real and imaginary parts of the energy eigenvalues $E_1$ and $E_3$; keeping $E_2$ as an isolated state. For $\lambda_I = 0.1$ (top panel), $\Re(E)$ of two interacting levels anti-cross and $\Im(E)$ cross; whereas, for $\lambda_I = 0.2$ (bottom panel) vice versa. The parameters are chosen as $\varepsilon_1 = 1$, $\varepsilon_2 = 1.5$, $\varepsilon_3 = 2$, $\omega_1 = 1$, $\omega_2 = -0.2$, $\omega_3 = -1$ and $\xi = 0.2$.}
	\label{figure_2}
\end{figure}

This matrix form can be trivially extended for specific higher dimensional applications. Here, $\lambda$ is a real/ complex tunable constant and an unitary transformation across the parameter $\lambda$ is executed by the matrix $U$, parametrized by $\xi$. The form of the unitary matrix $U$ is intentionally chosen to explore the proposed coupling scheme only. In the perturbation part, $\omega_i (i = 1,2,3)$, represent the coupling terms. Now the eigenvalues of $H$ are given by
\begin{subequations}
	\begin{align}
	& E_{1,3}(\lambda)=\frac{\varepsilon_1+\varepsilon_3+\lambda\left(\omega_1+\omega_3\right)}{2}\pm {C}\\	
	& E_{2}(\lambda)=\varepsilon_2+\lambda\omega_2
	\end{align}
	\label{equation_5} 
\end{subequations}
where,
\begin{equation}
{C}=\biggl[\left(\frac{\varepsilon_1-\varepsilon_3}{2}\right)^2+\left(\frac{\lambda\left(\omega_1-\omega_3\right)}{2}\right)^2
+\frac{\lambda}{2}\left(\varepsilon_1-\varepsilon_3\right)\left(\omega_1-\omega_3\right)\cos(2\xi)\biggr]^{1/2}
\label{equation_6} 
\end{equation} 
From the eigenvalue expressions (Eqns.~\ref{equation_5}), it is clearly observed that the coupling term $C$ (given by (Eqn.~\ref{equation_6})) appears only in the expressions of $E_1$ and $E_3$ but not in the expression of $E_2$. i.e. the states $E_1$ and $E_3$ are being interacted keeping $E_2$ as an isolated state. Clearly, at $\xi = 0$ two interacting levels $E_1$ and $E_3$ are degenerate at the point $\lambda = -(\varepsilon_1-\varepsilon_3)/(\omega_1-\omega_3)$. To lift this degeneracy one need to couple them by switching on $\xi$ and then avoided resonance crossing ($ARC$) occur between $E_1$ and $E_3$ with variation of the parameter $\lambda$. Now, to explore $EP$ with pertinent connection to this phenomena of $ARC$, the parameter $\lambda$ is chosen as a complex variable with form $\lambda = \lambda_R + i\lambda_I$. For a specific set of parameters all the three eigenvalues are plotted in Fig.~\ref{figure_2} as a function of $\lambda_R$ for two distinct values of $\lambda_I$. Interestingly, two different behavior of $ARC$s between $E_1$ and $E_3$ are clearly observed with anti-crossing and crossing between $\Re(E)$ and $\Im(E)$ in the top panel; whereas, crossing and anti-crossing between $\Re(E)$ and $\Im(E)$ in the bottom panel of Fig.~\ref{figure_2} respectively.
Accordingly, it is also observed that the change in $\lambda_I$ does not effect the intermediate state $E_2$ i.e. it remains isolated and varies as a function of $\lambda_R$ only. So in complex $\lambda$ plane two types of $ARC$s between $E_1$ and $E_3$ must be connected by a square root branch point singularity, whereas the intermediate state $E_2$ must be unaffected by such type of singularities. At this singular point (i.e. essentially where $C$ should be vanished) two interacting levels are being coalesced. Such singular point are called hidden singularity namely exceptional point ($EP$). So the $EP$ of the defined Hamiltonian is situated at a complex conjugate point given by
\begin{equation}
{\lambda_{EP}}=-\frac{\varepsilon_1-\varepsilon_3}{\omega_1-\omega_3}\exp(\pm 2i\xi)
\label{equation_7} 
\end{equation}
Now, the coupled energy eigenvalues can be expressed in terms of the characteristics of identified $EP$ as
\begin{equation}
E_{1,3}(\lambda)=E_{EP} \pm c_1\sqrt{\lambda-\lambda_{EP}}
\label{equation_8} 
\end{equation} 
In complex $\lambda$-plane, the values of $\sqrt{\lambda-\lambda_{EP}}$ on two different Riemann surfaces specify two distinct coupled energy levels, where the cross-joint of them represents the approximate $EP$ location. In similar way without losing any generality for higher dimensional situation, introducing a Hamiltonian matrix of the order $3n \times 3n$ $(n \in I)$ with appropriate coupling elements we should be able to establish the existence of such multiple second order $EP$s validating the proposed robust interaction restriction.

\section{Design of the Fabry-Perot type microcavity}

\subsection{Cavity specifications with operating parameters}

\begin{figure}[t]
	\centering
	\includegraphics[width=8.5cm]{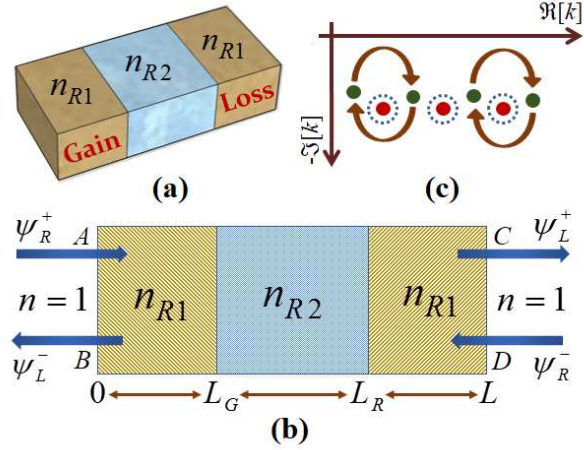}
	\vspace{0.2 cm}
	\caption{\textbf{(a)} (Color online) $3D$ schematic diagram of the designed Fabry-Perot type optical microcavity with nonuniform back ground refractive index; \textbf{(b)} $2D$ cross-sectional view of the same microcavity occupying the region $0\le x\le L$ with $L=10$ $\mu m$. Here $L_G=3$ $\mu m$ and $L_R=7$ $\mu m$ . The real background refractive indices are as $n_{R1}=1.5$ and $n_{R2}=4.5$. The eigenstates $\psi_L^+$ and $\psi_R^-$ indicate the incident waves with complex amplitudes $A$ and $D$ whereas the eigenstates $\psi_L^-$ and $\psi_R^+$ indicate the scattered waves with complex amplitudes $B$ and $C$ respectively; \textbf{(c)} Schematic non-linear distribution of $S$-matrix poles in complex $k$-plane of the microcavity under operating condition. The poles indicated by green circles represent the pair of interacted states whereas the poles indicated by red circles represent the isolated states.  }
	\label{figure_3}
\end{figure}

In order to achieve our goals, we design a two port Fabry-Perot type open optical microcavity with one dimensionally nonuniform background refractive index i.e. $n_R(x)$, as schematically shown ($3D$ view) in Fig.~\ref{figure_3}(a). Fig.~\ref{figure_3}(b) represents the $2D$ cross-section of the same microcavity which occupies the region $0\le x\le L$. Along length scale the distribution of $n_R(x)$ is given as follows.
\begin{equation}
n_R(x)=
\left\{ 
\begin{array}{ll}
n_{R1}, & 0\le x\le L_G\\
n_{R2}, & L_G\le x\le L_R\\
n_{R1}, & L_R\le x\le L\\
\end{array}
\right.
\label{equation_9} 
\end{equation} 
Now to add non-hermiticity, the cavity is pumped partially by introducing spatially unbalanced gain (with co-efficient $\gamma$) and loss profile in the two halves i.e. in the regions $0\le x\le L_G$ and $L_R\le x\le L$ respectively maintaining a fixed loss-to-gain ratio $\tau$. Hence, for $\gamma=0$ the cavity behaves like an Hermitian system. Under operating condition the refractive indices of the gain and loss regions are denoted by $n_G$ and $n_L$ respectively which can be expressed as
\begin{subequations}
	\begin{align}
	n_G & = n_R-i\gamma,\quad 0\le x\le L_G\\
	n_L & =  n_R+i\tau\gamma,\quad L_R\le x\le L
	\end{align}
	\label{equation_10} 
\end{subequations}
Specifically, for $\gamma \not=0$ the parameter $\tau$ can adjust the incorporated non-hermiticity independently in terms of system openness and coupling strength. For a fixed value of $\tau=1$, $\mathcal{PT}$-symmetry is conserved. But during operation, we set the parameter $\tau \not= 1$ to avoid $\mathcal{PT}$-symmetry constraint deliberately. 

\subsection{Scattering matrix formalism for calculation of eigenvalues}

Numerically, to study the resonance interaction phenomena in the designed microcavity we adopt a established method of scattering matrix ($S$-matrix) formalism where virtual states of resonances of the Hamiltonian corresponding to the real system are calculated in terms of poles of the associated $S$-matrix~\cite{Laha2,Laha1}. Using electro-magnetic scattering theory, here the matrix elements are analytically calculated as function of real system parameters. Now, for the designed cavity associated $S$-matrix can be defined through the input and output eigenstates relation given by
\begin{equation}
\left(\begin{array}{c}B\\C\end {array}\right)=S(n(x),\omega)\left(\begin{array}{c}A\\D\end {array}\right)
\label{equation_11}
\end{equation} 
Exploiting numerical root finding method, the poles of the defined $S$-matrix are calculated by solving the equation
\begin{equation}
\frac{1}{max[eig S(\omega)]}=0
\label{equation_12}
\end{equation}
Here, the denominator in L.H.S. of Eqn.~\ref{equation_12} gives the maximal-modulus eigenvalues of the matrix $S(\omega)$. 

Obeying current conservation and causality conditions, the $S$-matrix poles are calculated only at the lower half of the complex frequency plane ($k$-plane) for physical acceptability. Distribution of the calculated poles are schematically shown in Fig.~\ref{figure_3}(c). Interestingly, a nonlinear pattern have been observed in the pole distribution. Contextually, the equidistant linear distribution of $S$-matrix poles have already shown by choosing a same form of Fabry-Perot type optical microcavity with uniform background refractive index along length scale~\cite{Laha2}. But to establish the proposed next nearest neighbor coupling scheme, such nonlinearity in pole distribution is deliberately introduced by choosing nonuniform background refractive index along cavity-length, where only tuning such non-uniformity the distribution in $S$-matrix poles may be controlled as required for specific purposes. Associated interaction phenomena between the matrix poles (as shown in Fig.~\ref{figure_3}(c)) are topologically controlled through the spatial variation of unbalanced gain-loss profile with tunable parameters $\gamma$ and $\tau$. Accordingly, three pairs of interacting poles are deliberately identified keeping an intermediate isolated pole between each pairs. During operations, cavity is accompanied by avoided crossings between the interacting $S$-matrix poles.

\section{Numerical results towards encounter of hidden singularities with associated optical performances}

\subsection{Identifying the hidden singular points}

\begin{figure}[t]
	\centering
	\includegraphics[width=13cm]{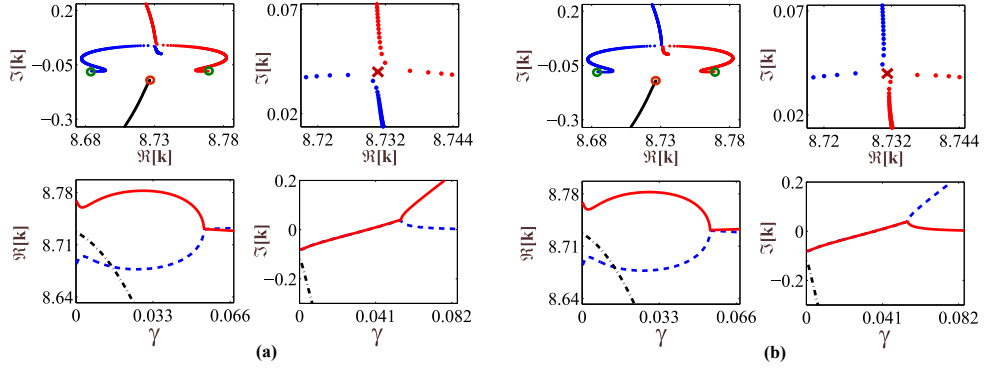}
	\vspace{0.2cm}
	\caption{(Color online) Trajectories of three chosen poles (dotted blue, red and black line) exhibiting $ARC$s (clearly shown in upper panel for both \textbf{(a)} and \textbf{(b)}) followed by the partial pumping in terms of unbalanced spatial gain-loss distribution in the cavity. In the passive cavity two green circles indicate the position of two poles which are interacting and the red circle denotes the position of intermediate isolated pole. The loss-to-gain ration is set at $\tau = 5.32$ in \textbf{(a)} and $\tau = 5.33$ in \textbf{(b)} respectively. The crossing/ anticrossing behavior of $\Re(k)$ and $\Im(k)$ are separately depicted as a function of $\gamma$ for both the $\tau$ values in lower panels. The red crosses in the top right panels represent the approximate positions of branch point singularities.}
	\label{figure_4}
\end{figure}

To encounter an $EP$, the mathematical concept of $ARC$s between eigenvalues of the matrix Hamiltonian (Eqn.~\ref{equation_3}) as delineated in section~\ref{matrix_formulation}, has been exploited where the cavity resonances (i.e. the eigenvalues) are treated as associated $S$-matrix poles. Accordingly, in the passive cavity three distinguish poles are precisely chosen over a particular frequency range. Then introducing the spatially unbalanced gain-loss profile by tuning the parameters $\gamma$ and $\tau$ the chosen poles are forced to interact mutually. It has been observed that with introduction of non-uniform gain-loss in the cavity, a pole belonging to the chosen set is being coupled with the pole situated at the next-nearest-neighbor position, whereas the intermediate pole remains unaffected. Now, for two distinct values of $\tau$, the evolution of resonance energies and widths with increasing amount of the parameter $\gamma$ are plotted in Fig.~\ref{figure_4} in terms of the $\Re(k)$ and $\Im(k)$ (i.e. real and imaginary part of the frequencies) of the chosen S-matrix poles. At first, we have set the parameter $\tau=5.32$ and accordingly slowly tuned the gain-coefficient $\gamma$ from $0$ to $0.1$. In this situation the trajectories of $S$-matrix poles are depicted in Fig.~\ref{figure_4}(a). The level repulsion phenomenon between the poles appearing in the next neighbor position are clearly observed in the upper panel with a zoomed in view. With increase in $\gamma$, the $\Re(k)$ experiences crossing whereas the $\Im(k)$ undergoes anti-crossing as shown in the lower panel. But for slight increase in $\tau=5.33$, a different behavior of level repulsion phenomenon has occurred as shown in Fig.~\ref{figure_4}(b); where the exchange in identities between the coupled poles (i.e. change in evolution direction from the previous case) have clearly observed in the upper panel. In this situation $ARC$ occurs with $\Re(k)$ undergoing anti-crossing and $\Im(k)$ experiencing crossing as shown in lower panel. Now, in both cases it is clearly observed that the intermediate pole is not effected by the incorporated non-hermiticity in the resonator. Even in significant change of the parameter $\tau$ it remains unaffected by other coupled poles and behaves as an isolated pole with change in the parameter $\gamma$.

Thus the behavior of $ARC$s between the coupled poles for two different $\tau$ values as depicted in Fig.~\ref{figure_4}(a) and~\ref{figure_4}(b) are topologically dissimilar. So, there must be an abrupt transition between two $\tau$-values where the coupled poles coalesce at a critical square root singular point in $(\gamma, \tau)$-plane; at which the associated eigenfunctions loose their identities. In complex $k$-plane, the positions approximately indicated by red crosses in upper panel (right side) of both Fig.~\ref{figure_4}(a) and~\ref{figure_4}(b) respectively are identified as the appearance of such singular point. For our chosen specific set of cavity parameters, in $(\gamma, \tau)$-plane the position of this singular point have found at $\sim(0.055, 5.328)$.

\subsection{Cascaded state flipping around the identified singularity}

\begin{figure}[t]
	\centering
	\includegraphics[width=8.5cm]{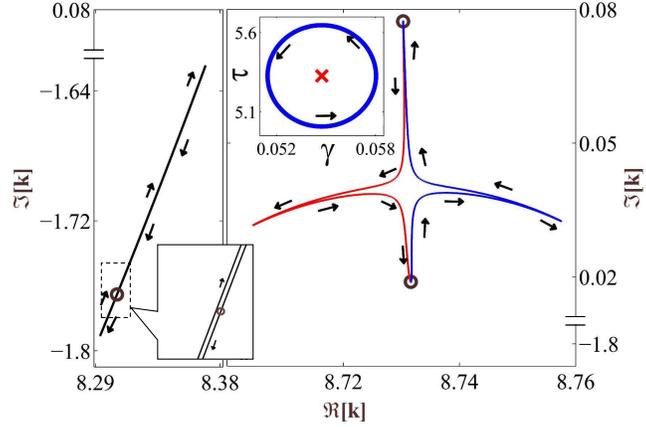}
	\vspace{0.2cm}
	\caption{(Color online) Trajectories of the three poles in complex $k$-plane (initial positions are marked by the brown circles) associated with the identified singularity denoted by red cross in $(\gamma, \tau)$-plane at inset for an encircling process (blue circular path at inset) centering it with $a=0.04\ a.u.$. In $k$-plane, dotted red and blue lines represent the trajectories of the coupled poles whereas dotted black line represent the trajectories of the intermediate isolated poles after one round encirclement around the singularity in $(\gamma, \tau)$-plane. The dynamics of coupled poles are depicted with respect to the $\Im(k)$ axis where the ticks labels are shown in the right sides, while the left side distribution in ticks labels correspond to the $\Im(k)$ axis to depict the dynamics of isolated pole. Such two different distribution in ticks labels corresponding to the $\Im(k)$ axis is considered for clear visibility. A zoomed in view around the passive position of the intermediate pole is also shown for clear visibility in loop formation.}
	\label{figure_5}
\end{figure}
In this part we have explored an unique feature of the designed microcavity in the vicinity of the identified hidden singularity towards flipping of cavity resonances in the context of optical mode converter. Accordingly, we study the effect of encircling around this singular point. We choose circle with center at $(\gamma_0, \tau_0)$ as a closed loop in $2D$~($\gamma, \tau$)-plane which can be expressed by the following parametric equation~\cite{Cartarius3,Laha2,Ghosh,Menke}
\begin{subequations}
	\begin{align}
	&\gamma(\phi)=\gamma_{0}\left[1+a\,cos(\phi)\right]\\
	&\tau(\phi)=\tau_{0}\left[1+a\,sin(\phi)\right]
	\end{align}
	\label{equation_13} 
\end{subequations}
where, $a$ $(\in [0,1])$ represents a certain small characteristics parameter (equivalent to radius of the circle) and $\phi$ $(\in [0, 2\pi])$ is a tunable angle. This method opens up a possibility to scan a large area around the singularity at once. Choosing enough small steps on the enclosing loop, motion of resonances can be properly traced. Once the described circle in parameter place is traced, successively the position exchanging behavior between the eigenvalues gives the proof about existence of an exceptional point~\cite{Cartarius3,Laha2,Ghosh,Menke}.   

Now choosing the identified singularity as the center, a closed contour is patterned in ($\gamma, \tau$)-plane with $a=0.04$ $a.u.$ as shown at the inset of Fig.~\ref{figure_5}. The parameter $a$ have chosen in such a way that the described contour should rightly enclose the identified singularity. An anticlockwise operation has been performed along this closed contour. Now Fig.~\ref{figure_5} shows how the coupled pair of poles (appear in next nearest neighbor positions) are associated with the singularity and the intermediate isolated pole is affected by such encircling process in complex $k$-plane. Here, in $k$-plane each point on the red blue and black trajectories indicate the point-to-point evolution of $S$-matrix poles from their starting positions (represented by the brown circles) with associated encircling process (following green circle at the inset) around the respective singularity (denoted by red cross at the inset) in ($\gamma, \tau$)-plane. Interestingly, as the result of one round encircling process in parameter plane around the singularity, two coupled poles have exchanged their positions in complex $k$-plane. However, the intermediate pole has completed an individual loop in same plane i.e. after complete of encircling process it returns to its initial position. Thus the intermediate pole remains unaffected by the presence of singularity inside the enclosing loop. Accordingly, another encirclement around the singularity (i.e. total two successive rounds) results that the pair of coupled poles regain their initial positions with formation of a complete loop; whereas, intermediate pole makes an extra loop exactly along the previous path in $k$-plane. Such position exchanging behavior between the coupled poles may be called as the \textit{flip-of-states}. Trajectories marked by dotted red, blue and black curves are associated with the pair of coupled poles and intermediate isolated pole respectively. Arrows indicate the direction of progression in both $k$-plane and ($\gamma, \tau$)-plane.

Now, we demonstrate a numerical observation behind the reason for isolation of intermediate pole even in presence of sufficiently effective pumping in terms of spatially unbalanced gain-loss profile. The complex cavity resonances are listed in the table~\ref{table_1} for both passive and initial pumped conditions.
\begin{table}[h]
	\caption{Complex resonances of the microcavity for both passive and initial pumped conditions. All values are given in absolute unit.}
	\centering
	\small
	\begin{tabular}{|c||l|l||l|l|}
		\hline
		\multirow{2}{*}{\thead{States associated\\ with the singularity}} 
		& \multicolumn{2}{c||}{\thead{Passive cavity\\ resonances}}
		& \multicolumn{2}{c|}{\thead{Initially \\ pumped cavity \\ resonances}} \\             \cline{2-5}
		& $\Re(k)$ & $\Im(k)$ & $\Re(k)$ & $\Im(k)$ \\  
		\hline
		State-1 & 8.683 & - 0.078 & 8.731 & 0.021  \\      \hline
		State-2 (intermediate) & 8.726 & - 0.118 & 8.306 & - 1.765  \\      \hline
		State-3 & 8.770 & - 0.078 & 8.730 & 0.068  \\		\hline
	\end{tabular}
	\label{table_1}
\end{table}
It has been observed that in passive cavity the imaginary part of the resonance frequencies ($\Im(k)$) are almost equal for two poles appear in next nearest neighbor position whereas the intermediate pole appears at a lower $\Im(k)$. Now when the encircling process around the singularity starts, certain amount of gain and proportionate loss (which depends on position of the respective singularity in ($\gamma, \tau$)-plane and the characteristics parameter $a$ of the enclosing loop) is imposed instantly on each of the three poles; which results in starting their movement. So after this instant initial pumping, it has been significantly noticed that there is a huge change in the value of $\Im(k)$ of the poles appearing in the next nearest neighbor positions, however there is a very small change in $\Im(k)$ of the intermediate pole. Due to this anamolous behavior of poles in $\Im(k)$, we display results in Fig.~\ref{figure_5} in a different manner for clear visibility; where in same $k$-plane the trajectories of two coupled pole and the intermediate pole are depicted with two different distribution in $\Im(k)$-axis as shown in the right and left sides respectively. For clear visibility in the loop formation for the case of intermediate pole, we also present a zoomed out view around its position in $k$-plane. Essentially, the factor $\Im(k)$ physically represents the resonance width. Thus from this numerical observation it can be said that the resonance width may be responsible for this anamolous interaction process~\cite{Menke}. Because of sufficient change in $\Im(k)$, the poles appearing in the next nearest neighbor positions are coupled as we increase the pumping, and depicts the phenomena of flip-of-sates following the adiabatic encirclement around associated singularity. Whereas due to exact opposite behavior of $Im(k)$ (i.e. very little change due to initial pumping) of the intermediate state, it remains unaffected by the presence of singularity even in addition of sufficient pumping.
\begin{figure}[t]
	\centering
	\includegraphics[width=7cm]{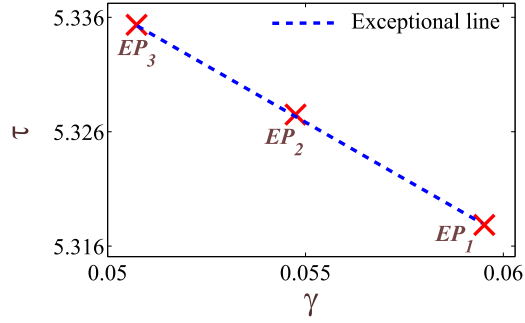}
	\vspace{0.2 cm}
	\caption{Approximate locations of three embedded $EP$s denoted by three red crosses in the ($\gamma, \tau$)-plane. Blue dashed line represents the exceptional line with negative tangent.}
	\label{figure_6}
\end{figure}
\begin{figure}[t]
	\centering
	\includegraphics[width=13cm]{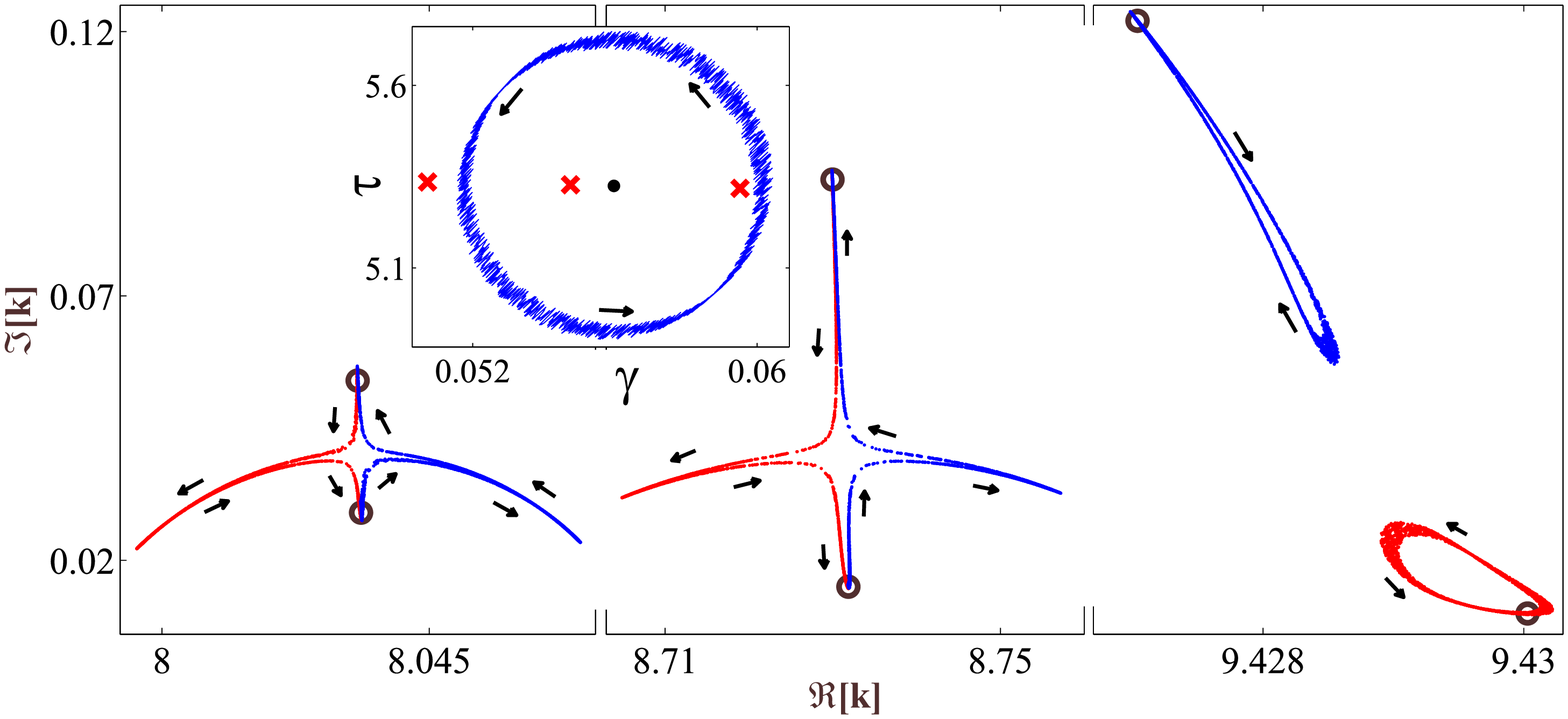}
	\vspace{0.2 cm}
	\caption{(Color online) Trajectories of the coupled eigenvalues (initial positions are marked as the brown circles) corresponding to all three consecutive $EP$s in complex $k$-plane (denoted as three red crosses at the inset) for a common encircling process in $(\gamma, \tau)$-plane around the center at $\sim(0.056, 5.325)$ (marked as blue dot in inset) with $a=0.065\ a.u.$. in presence of modest random fluctuations on the enclosing contour (described as blue fluctuated circular path at inset). Here the described prametric path encloses both $EP_1$ and $EP_2$, except $EP_3$.}
	\label{figure_7}
\end{figure}

Thus the state flipping behavior between the pair of coupled poles associated with the identified singularity clearly establishes the fact that their dynamics is entirely controlled by the exception point ($EP$) even in presence of intermediate isolated pole. Here the $EP$ conventionally exhibits as a second order branch point for eigenvalues. This is the direct observation of exceptional singular behavior of a hidden branch point~\cite{Cartarius3,Laha2,Ghosh,Menke}.                          

\subsection{Formation of hidden singular line}

Similarly, tuning the factor $\tau$ over the amount of gain-coefficient $\gamma$, we have encountered at least three different $EP$s in $(\gamma,\tau)$-plane with deliberate identification of three different set of poles, where each set must contain three distinguished poles according to the proposed scheme. To correlate all the identified $EP$s, we plot them in $(\gamma, \tau)$-plane as shown in Fig.~\ref{figure_6} where each red cross denotes each of the embedded $EP$s in the cavity. Here, blue dotted line gives the best fitting which indicates that the identified $EP$s follow a special straight line in parameter plane which may be called as hidden singular line. Here such hidden singular line is termed as {\it exceptional line}. Previously, this formation of exceptional line has been reported for the first time by the authors while considering the nearest neighbor coupling situations between the consecutive poles to explore $EP$s in a different class of Fabry-Perot microcavities~\cite{Laha2}. As all the identified second order $EP$s in the designed cavity are correlated by an single exceptional line, successive state switching between the $EP$s has been achieved straight forwardly i.e. simply by tuning the coupling parameters $\gamma$ and $\tau$ adiabatically. Towards the exploration of unconventional optical effects associated with $EP$s, the formation of such exceptional line brings in a new degree of freedom for manipulation of cavity resonances.

\subsection{Stable optical performance towards cascaded state-flipping mechanism}

Exploiting the special feature of exceptional line, we explore the stable optical performance of the designed microcavity in the vicinity of identified $EP$s via numerical exemplification of robustness of associated flip-of-state phenomena. A common encircling process in parameter plane has been chosen to explore such flip-of-state phenomena corresponding to all the identified $EP$s in complex $k$-plane at a time. Associated results are displayed in Fig.~\ref{figure_7}. To consider all the $EP$s with respect to the same enclosing process, the center of described loop has chosen at any arbitrary point following exceptional line in ($\gamma, \tau$)-plane say at $\sim(0.056, 5.325)$ (denoted by black dot at the inset); i.e. none of the $EP$s represent the center of described contour in $(\gamma, \tau)$-plane. The characteristics parameter $a$ has chosen as $0.075$ $a.u.$ to enclose $EP_1$ and $EP_2$ only; except $EP_3$ (as shown in inset of Fig.~\ref{figure_7}). So we can observe the effect of encircling on $EP$s for both inside and outside the enclosing loop in same $k$-plane. Here exceptional line has been exploited to shift the encircling parameter set around the particularly chosen arbitrary center between three identified $EP$s and accordingly the dynamics of the poles corresponding to each $EP$ have been analyzed. We have also added some deliberate random fluctuations on the enclosing loop to substantiate the rigidity of described state-flipping behavior associated with each $EP$ against unwanted fabrication tolerances of state-of-the-art techniques due to various real natural effects.

Now following the one round encircling process along the described contour in $(\gamma, \tau)$-plane, it has been observed that the coupled poles associated with $EP_1$ and $EP_2$ are exchanging their positions in a very generic fashion; while the coupled poles corresponding to $EP_3$ are constructing the individual loops in complex $k$-plane. Accordingly, for two successive encirclement along the contour in parameter plane the eigenvalues corresponding to the first two $EP$s have formed complete loop (two individual complete loop corresponding to different $EP$) in $k$-plane after second permutation; whereas eigenvalues corresponding to $EP_3$ traversed the exact previous path once again avoiding any kind of permutations. We also study the trajectories of intermediate isolated poles associated with each $EP$. However, here also they have behaved in previous manner as expected i.e. all of them are unaffected by presence of other singularities and remain isolated making individual loops followed by the described encirclement process. Hence purposely we exclude the trajectories of isolated poles form Fig.~\ref{figure_7} for clear visibility of trajectories of the coupled poles associate with the described encirclement process around the $EP$s. Thus the flip-of-states phenomena around $EP$s is omnipresent until the one-to-one coupling restriction is topologically preserved and the identified $EP$s must be untouched by the deformations in parametric loop. Contextually, due to substantial modifications in the encircling process the microcavity may support secondary states unconventionally which can interact with both scattering as well as isolated states in the microcavity~\cite{Laha2}. As a result, described one-to-one coupling restriction between the resonances with an intermediate isolated resonance may be interrupted and then the singular behavior of the identified $EP$s may be destroyed i.e. state-flipping behavior should no longer be stabled. 

Thus from the results described in Fig.~\ref{figure_7} the following conclusions should be drawn. The state-flipping behavior in $k$-plane corresponding to each identified $EP$ is independent as unaffected by the presence of any other $EP$ inside or outside the contour in $(\gamma, \tau)$-plane. Such optical performances of the designed cavity present the robust behavior even in presence of parameter fluctuations/ deformations. Hence, robust behavior of flip-of-state phenomena is extremely promising for device level implementation using any state-of-the-art fabrication technique with modest tolerances.

\section{Conclusions}

In summary, using $S$-matrix formalism we have modeled a non $\mathcal{PT}$-symmetric two port open Fabry-Perot styled optical microcavity to explore a non-trivial next nearest neighbor resonance interaction which is entirely controlled by the system topology i.e. geometry of the cavity and spatial distribution of unbalanced gain-loss. Non-uniform variation in background real refractive index has adopted purposely  to introduce an inherent non-linearity in distribution of $S$-matrix poles in complex eigenvalue plane. Adjusting the factor $\tau$ over gain co-efficient $\gamma$, three different second order $EP$s have been embedded in operating parameter plane of the cavity under strict restriction of one-to-one coupling. Unveiling the formation of special exceptional line in the parameter plane supported by each identified $EP$s, we explore unique cascaded state-flipping mechanism between the coupled poles corresponding to the encircled $EP$s with successive state switching between them along the reported special line. We have established that if an $EP$ is rightly encircled by the parametric loop which may centered either at that $EP$ or any arbitrary point following the exceptional line, then associated flip-of-state phenomena is ubiquitous. Moreover, this occurs irrespective from presence of any other $EP$ inside the described parametric loop. Such state-flipping mechanism is evident even in presence of moderate deformation/ fluctuations on parameter variation during encircling process. Hence exploring the special next-nearest-neighbor coupling scheme and exploiting the concept of exceptional line, stable optical performance of such degenerate microcavity has been achieved by establishing the robustness of unique state-flipping behavior in the vicinity of $EP$s even in presence of an intermediate isolated resonance between two coupled resonances. Recent developments in the fabrication technology for growth of such optical microcavities with high precision and control may open up a new platform to implement high performances integrated photonic devices on chip.

\section*{Funding Information}

This work acknowledges the financial support by Department of Science and Technology (DST), India under the INSPIRE Faculty Fellow grant [IFA-12; PH-23].

%Manual citation list

\end{document}